\documentclass[runningheads]{llncs}
\usepackage{lmodern}
\usepackage{textcomp}
\usepackage{newtxtext,newtxmath}
\usepackage{graphicx}
\usepackage{comment}
\usepackage{multirow}
\usepackage{xcolor}
\usepackage{hyperref}

\usepackage{csquotes}
\usepackage{balance}

\usepackage{algorithm}
\usepackage{algorithmic}
\usepackage{amsmath}
\usepackage{cleveref}
\usepackage{amsfonts}

\begin{document}

\newcommand\submittedtext{%
\textbf{Preprint}. This work has been accepted to the 30$^{th}$ International Conference on Enterprise Design, Operations, and Computing (EDOC 2026). Personal use of this material is permitted. Permission must be obtained for all other uses, in any current or future media.}

\newcommand\submittednotice{%
\begingroup
\renewcommand\thefootnote{}%
\footnotetext{%
\noindent\fbox{\parbox{\dimexpr\linewidth-2\fboxsep-2\fboxrule\relax}{\submittedtext}}%
}%
\endgroup
}

\title{An Evidence-driven Protocol for Trustworthy CI Pipelines}%

\titlerunning{Evidence-driven Protocol for TCI  Pipelines}

\author{
Fernando Castillo\inst{1} \and
Eduardo Brito\inst{2,3} \and
Pille Pullonen-Raudvere\inst{2} \and 
Sebastian Werner\inst{1} \and 
Stefan Tai\inst{1}
}

\authorrunning{F. Castillo et al.}

\institute{Information Systems Engineering, TU Berlin, Einsteinufer 17, Germany\\
\email{\{fc,sw,st\}@ise.tu-berlin.de} \and
Cybernetica AS, Estonia 
M\"aealuse 2/1, 12618 Tallinn, Estonia\\
\email{\{eduardo.brito, pille.pullonen-raudvere\}@cyber.ee}
\and
University of Tartu, Narva mnt 18, 51009, Tartu, Estonia
}

\maketitle
\submittednotice

\begin{abstract}

Enterprise software supply chains are increasingly vulnerable to infrastructure attacks, resulting in financial and reputational damage.
Ensuring the integrity and provenance of software artifacts remains a significant challenge, where re-execution of the build and tests by every consumer to guarantee provenance produces a verification bottleneck and credibility reduction.
This paper presents an evidence-driven protocol for trustworthy Continuous Integration (CI) pipelines that combines Deterministic Build Systems (DBS) with Trusted Execution Environments (TEEs). The approach provides cryptographically verifiable guarantees of integrity, authenticity, and attestation for CI artifacts in distributed environments, reducing implicit trust without requiring costly re-execution by consumers. We introduce a protocol that binds deterministic builds with TEE-based attestations, formalizing the evidence life cycle, together with a practical implementation using Nix and Intel TDX.
Experimental results show that artifact verification is reduced from redundant computation to lightweight signature and policy checks. These findings demonstrate that evidence-driven CI pipelines establish scalable and verifiable trust in digital infrastructure, effectively amortizing the initial computational overhead introduced by TEEs.

\keywords{Continuous integration security, Trustworthy software, Secure software supply chain, Evidence-driven trust
}

\end{abstract}

\section{Introduction}\label{ch:introduction}
Attacks on digital infrastructure, whether within enterprises or in external systems they depend on, can directly and indirectly compromise organizations and cause large-scale operational disruptions. In such scenario, software supply chain security has emerged as a critical area of research, particularly in response to high-profile attacks like SolarWinds~\cite{alkhadra2021solar}, Log4Shell~\cite{everson2022log4shell}, or the XZ exploit~\cite{lins2024critical}, which demonstrated the vulnerabilities inherent to modern software supply chains. 
Such incidents have caused losses of hundreds of millions of dollars, frequently exploiting vulnerabilities in the software supply chain of third-party components~\cite{williams2025research}, and most recently exemplified by the December 2025 Trust Wallet pipeline bypass causing \$8.5M USD in losses\footnote{\href{https://trustwallet.com/blog/announcements/trust-wallet-browser-extension-v268-incident-community-update}{Trust Wallet v2.68 Incident Update}}. %

As software development becomes increasingly complex and globally distributed, establishing trust in produced software artifacts has become progressively more difficult. For enterprise infrastructure, where code directly governs economic behavior, trustworthy artifacts must provide not only integrity, but also authenticity and attestation guarantees across the entire software life cycle. However, modern DevOps and CI/CD practices, while enabling rapid delivery, significantly hinder the enforcement of these guarantees~\cite{rajapakse2022challenges}. Security-sensitive steps such as building and testing are frequently outsourced to shared CI environments, expanding the attack surface of the software supply chain.

A fundamental weakness of current software supply chains is that deployment environments rarely rebuild software from source, instead relying on pre-built binaries or container images obtained from external infrastructure~\cite{fleischer2020memory}. Ensuring end-to-end trust from source commits to build artifacts and test results therefore remains an open challenge, which is exacerbated by growing codebases, extensive third-party dependencies~\cite{ishgair2024sok}, globally distributed teams, or development outsourcing~\cite{niazi2016challenges}. While approaches such as DevSecOps~\cite{sanchez2020security,britotrustops}, Software Bills of Materials (SBOMs)~\cite{xia2023empirical}, and Continuous Compliance~\cite{fitzgerald2017continuous} improve visibility and security practices, they lack scalable mechanisms to provide verifiable trust guarantees across heterogeneous and distributed execution environments. This reliance on implicit trust fuels a 'credibility deficit'~\cite{leblanc2025rethinking}, where users cannot confidently adopt artifacts without performing redundant validations.

In this paper, we introduce an evidence-driven protocol for trustworthy Continuous Integration (CI) pipelines that mitigates reliance on implicit trust assumptions. The protocol formalizes declarative, inspectable processes backed by cryptographic guarantees of integrity, authenticity, and attestation throughout the software supply chain. %
Our main contributions are as follows:
\begin{itemize}
\item \textbf{Novel Evidence-driven Trustworthy CI Protocol:} A generalized protocol that seamlessly integrates into standard CI pipelines, transforming implicit trust into a verifiable chain of evidence and enabling a lightweight artifact verification model without redundant rebuilds.%
\item \textbf{Implementation and evaluation:} A practical proof of concept using GitLab, NixOS, and Intel TDX-based TEEs, together with an evaluation demonstrating cost savings from avoided rebuilds and tests.
\end{itemize}

The remainder of this paper is organized as follows: Section~\ref{ch:background} introduces background concepts; Section~\ref{ch:rw} reviews related work; Section~\ref{ch:requirements} defines trust model; Section~\ref{ch:architecture} presents the protocol; Section~\ref{ch:implementation} describes the implementation; Section~\ref{ch:evaluation} evaluates the system; Section~\ref{ch:discussion} discusses its implications; and Section~\ref{ch:conclusion} concludes.

\section{Background}\label{ch:background}

In this section, we first describe how TEEs can produce evidence, and discuss how trust can be established in CI pipelines.

\subsection{Evidence from Trusted Execution Environments}
Trusted Execution Environments (TEEs) are hardware-supported isolation mechanisms that protect code and data from external tampering and disclosure~\cite{munoz2023survey}. Programs executed within a TEE run in isolated or encrypted memory, ensuring computation integrity even in the presence of a potentially untrusted system operator.
TEEs support remote attestation, enabling external parties to verify the integrity of the enclave’s execution state and the authenticity of its outputs. This mechanism prevents malicious entities from impersonating them. Each TEE is provisioned with a hardware-backed identity key at manufacturing time, verifiable via a Public Key Infrastructure (PKI). This key is used to derive attestable identities for instantiated TEEs, enabling the generation of verifiable evidence of correct execution.

Modern platforms such as Intel TDX (Trusted Domain Extensions) and AMD SEV (Secure Encrypted Virtualization) support VM-based TEEs, where entire virtual machines are protected by hardware-enforced isolation. These platforms provide confidentiality and integrity guarantees for VM memory and execution state, making them well suited for executing security-critical CI tasks in a verifiable manner. By combining secure execution, identity verification, and remote attestation, TEEs enable the production of evidence required to create a verifiable chain of trust across CI phases. 

\subsection{Trust in Continuous Integration Pipelines}
As software pipelines grow in scale and complexity, ensuring trust and transparency has become a central concern. Frameworks such as DevSecOps~\cite{sanchez2020security} and VeriDevOps~\cite{enoiu2023veridevops} integrate security and verification mechanisms throughout the software life cycle, while TrustOps~\cite{britotrustops} emphasizes continuous evidence collection and validation across development and operation phases.
CI pipelines automate the integration of code changes, enabling frequent builds, testing, and deployment while maintaining a consistent deployable state. However, as CI infrastructures expand to support distributed teams and multi-cloud environments, they face increasing challenges related to security, integrity, and traceability~\cite{bajpai2022secure}.

Deterministic build systems (DBS) establish a crucial foundation for software supply chain security by enabling consistent, reproducible pipelines~\cite{burr2019software}. Nevertheless, reproducibility alone does not guarantee the security or correctness of the resulting artifacts~\cite{jamthagen2016exploiting}, necessitating the explicit declaration and enforcement of security policy checks throughout the CI pipeline. Consequently, combining the determinism of a DBS with the attestation of a TEE presents a promising avenue for producing the verifiable evidence needed to secure CI pipelines.

\section{Related Work}\label{ch:rw}
In this section, we position our work with respect to existing approaches. Despite substantial progress, current solutions still lack comprehensive, verifiable evidence spanning the full software development and deployment life cycle. Our work addresses these gaps with an evidence-driven protocol.

\newcommand{\related}[1]{\textbf{\textit{#1.}}}

\textbf{Software Supply Chain Security Challenges:}
Software supply chain security has become a major research focus following high-profile incidents such as ByBit~\cite{monperrus2025software}, SolarWinds~\cite{alkhadra2021solar}, and the XZ exploit~\cite{lins2024critical}, which exposed systemic vulnerabilities in modern supply chains and caused losses exceeding hundreds of millions of dollars. Tools such as Trivy and other SBOM-based solutions aim to improve transparency and traceability of software components~\cite{o2024assessing}, with recent work extending dependency analysis~\cite{liu2025dirty} and policy auditing or usage justification~\cite{stengele2025supply}. However, important gaps remain, including the lack of unified threat modeling across integration, CI, and CD phases~\cite{reichert2024software}, as well as inconsistencies between SBOM outputs of different tools~\cite{yu2024correctness}, which undermine reproducibility and trust.

\textbf{Limitations of DevSecOps:}
DevSecOps promotes embedding security controls throughout the software life cycle, including testing and vulnerability analysis. However, it does not provide a systematic approach for managing and verifying the evidence produced by these activities~\cite{britotrustops}, nor does it fully address trust guarantees required by frameworks such as SLSA~\cite{slsa2024supply}. For example, in~\cite{mahboob2021kubernetes}, TEEs are employed within CI pipelines, but without producing attestable guarantees for final artifacts, leaving trust assumptions confined to the execution environment rather than the end-to-end pipeline.

\textbf{Limitations of Existing Systems and Frameworks:}
Several systems improve specific aspects of supply chain security but fall short of comprehensive, verifiable evidence generation. VeriDevOps integrates automated policy verification into DevOps workflows~\cite{sadovykh2021veridevops}, but does not produce independently verifiable evidence. in-toto~\cite{torres2019toto} focuses on securing SBOM integrity across the supply chain, while related work on distributed builds emphasizes SBOM reproducibility using Merkle trees~\cite{lew2024distributed}. However, evaluations show that many SBOM tools do not meet minimum requirements defined by security authorities such as the U.S. National Telecommunications and Information Administration~\cite{halbritter2024accuracy}. Sigstore~\cite{newman2022sigstore} provides artifact signing through developer authentication, ephemeral keys, and transparency logs, but it does not address execution environment attestation or deterministic build guarantees. Its reliance on append-only logs and ephemeral keys also introduces exposure to denial-of-service and key-reuse attacks~\cite{newman2022sigstore}. Ultimately, without cryptographically verifiable evidence to enforce these security claims, existing frameworks fail to resolve the credibility deficit in modern software ecosystems~\cite{leblanc2025rethinking}.

In summary, these limitations highlight the need for an integrated approach that (i) enables continuous and verifiable evidence collection across the software life cycle, (ii) integrates deterministic execution guarantees directly into the CI pipeline, and (iii) provides guarantees of authenticity, integrity, and attestation in distributed software supply chains to replace redundant rebuilds with constant-time verification checks. %

\section{Trust Model and Trust Mechanisms}\label{ch:requirements}
In this section, we first describe the Trust Model we use as reference, based on known security practices and standards~\cite{chandramouli2024strategies,souppaya2022secure,slsa2024supply}, including the actors/entities involved and the trustworthiness requirements. Later, we introduce the mechanisms to address those requirements with the proposed evidence-driven protocol.

\subsection{Trust Model} \label{sec:trust_model}

We first define the entities (\textbf{E}) in the baseline CI pipeline:

    \textbf{E1 - Repository Owner:} holds the primary responsibility for defining security policies, including permissions, access controls, and workflow triggers and setting the rules governing the CI pipeline.
    
    \textbf{E2 - Developers:} write and sign commits in the repository. 
    
    \textbf{E3 - CI Administrators:} manage the CI infrastructure.
    
    \textbf{E4 - Version Control System (VCS):} stores and versions the source code in a repository. 
    
    \textbf{E5 - Artifact Registry:} stores the artifacts.
    
    \textbf{E6 - CI Pipeline Infrastructure:} where the code is built and build artifacts are tested.
    
    \textbf{E7 - Deployment Environment:} The final destination for the artifacts, e.g., package managers, production environment.%

\paragraph{}
We use the following scenarios (\textbf{S}) for possible threats, based on existing characterizations of CI pipeline security properties~\cite{koishybayev2022characterizing,chandramouli2024strategies}:%

\textbf{S1 - Compromised Integrity of Pipeline Processes:} Malicious modifications or unintended alterations in the build and test environments can lead to compromised artifacts, for example, from CI administrators or developers potentially introducing vulnerabilities or backdoors, through code or workflow modifications, e.g., the Bybit attack. %

\textbf{S2 - Dependency of Third-Party Components:} The use of third-party dependencies in the source repository introduces the risk of vulnerabilities or malicious code within these components, which could affect the final software product, e.g. the ReactJS 2025 vulnerability (CVE-2025-55182)\footnote{https://www.cve.org/CVERecord?id=CVE-2025-55182}. %

\textbf{S3 - Insufficiency of Continuous Evidence and Audit Trails:}  Insufficient evidence collection and traceability from the CI pipeline can compromise the ability to detect, investigate, or establish accountability if an incident occurs.%

\paragraph{}

To mitigate these threats and establish operational credibility, we derive the following trustworthiness requirements (\textbf{TR}) for artifacts produced in CI pipelines:

\textbf{TR1 - Source Integrity and Authenticity:} Ensure that source code and version control data originate from authorized and verified sources.
    
\textbf{TR2 - Environment Integrity:} Guarantee that build and test environments are isolated, immutable, and operate in a known, verifiable state.

\textbf{TR3 - Process Integrity:} Ensure that pipeline processes execute as defined, without unauthorized modifications or deviations.

\textbf{TR4 - Artifact and Evidence Integrity and Authenticity:} Ensure that CI artifacts and evidence generated during the pipeline are authentic, untampered, retaining their integrity throughout the CI processes.

\subsection{Trust Mechanisms}

\begin{figure*}[h]
    \centering
    \includegraphics[width=0.98\textwidth]{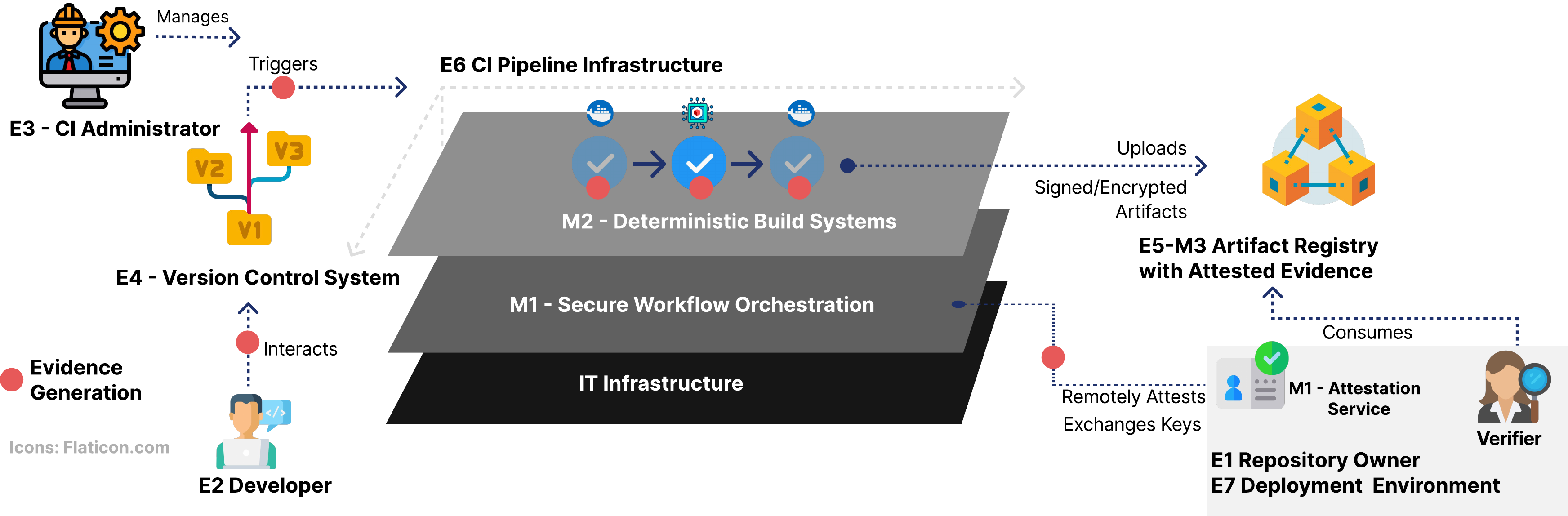}
    \caption{Diagram of Trust Mechanisms in a CI Pipeline }
    \label{fig:trustworthy-ci-architecture}
\end{figure*}

The following mechanisms \textbf{(M)} are intended to fulfill the trustworthiness requirements of the trust model by together producing the necessary evidence of the executed CI pipeline, illustrated in~\Cref{fig:trustworthy-ci-architecture}.%

\subsubsection{M1 - Secure Workflow and Orchestration Engines:}

Modern software development and deployment pipelines rely on complex workflows composed of interdependent tasks that execute in a predefined sequence. Workflow engines programmatically control these processes using a directed acyclic graph (DAG) structure, where nodes represent tasks and edges define dependencies. Each task produces and consumes evidence in the form of inputs and outputs, ensuring that its execution can be validated based on the successful completion of preceding tasks. %
Pipeline orchestration specifications define the tasks, their relationships, and metadata, including environment configurations, triggers, and runtime requirements, offering a high-level abstraction for creating and managing pipelines.
For trustworthy CI/CD pipelines, the orchestration layer plays a role in ensuring authenticity, integrity, and attestation. To support this, pipeline specifications must clearly identify and configure tasks that run within a TEE. %

To maintain the integrity, authenticity, and confidentiality of processes running within TEEs, a KMS is essential in the CI pipeline pipeline. The KMS manages cryptographic keys used for verifying the TEE’s identity, enabling remote attestation, and securely signing or encrypting CI outputs, such as build artifacts, test results, or audit logs. Proper key management ensures that only authorized keys are used, allowing the production of verifiable and, when necessary, confidential outputs as evidence that external verifiers can attest or decrypt.
Before any CI task can be executed in a TEE, the environment must undergo remote attestation to confirm its identity and integrity.

\subsubsection{M2 - Deterministic Build Systems:}

In a reproducible build, running the same process in the same environment should consistently produce the same results, regardless of when or where it is executed~\cite{shi2021experience}. This reproducibility is especially crucial in scenarios involving TEEs, where independently verifiable build processes are essential to establish trust in the integrity of the system~\cite{jamthagen2016exploiting}. %
A fully specified, reproducible build process enables authentication and distribution of the build specification as authenticated evidence. The specification can then be tracked, attested, and used within TEEs. In practice, a declarative build specification could be stored in a trusted registry as part of the CI pipeline. Whenever a build is triggered, the orchestration engine would use the attested specification to recreate the environment within the TEE, ensuring that the build is reproducible, isolated, and verifiable through its secure boot, isolation and attestation mechanisms~\cite{shepherd2024trusted}.

\subsubsection{M3 - Artifact Registry with Attested Evidence:}

Once the CI pipeline completes, the produced artifacts are pushed into a secure artifact registry that ensures their integrity, authenticity, and traceability. This registry supports signed and encrypted artifacts, holding and providing the cryptographic proofs of their provenance.
Blockchain commitments add an extra layer of security by providing immutable and tamper-proof, publicly auditable records of CI pipeline evidence. By storing cryptographic commitments (e.g., hashes of artifacts or test reports) in a distributed ledger, organizations can ensure the integrity and traceability of their pipelines. %
Nonetheless, the artifact registry represents the final, crucial step in a secure CI pipeline. 
Only artifacts accompanied by attested evidence, such as test or security audit reports, should be allowed to be deployed or processed further once verified against a defined policy, safeguarding the integrity of the software supply chain.
\newline
\newline
Together, these mechanisms enhance the security, traceability, and verifiability of the CI pipeline, establishing trustworthy evidence for the software development environment.%

\section{Evidence-driven Protocol for Trustworthy CI Pipelines}\label{ch:architecture}

The following section explains the evidence life cycle and how the protocol processes this data to produce a single, verifiable attestation for the entire CI pipeline.

\subsection{Trustworthiness through Evidence}
To ensure the trustworthiness requirements are met, we formalize the collection of evidence throughout the CI pipeline. 
In the context of CI pipelines, evidence refers to the collection of data and artifacts that document each phase of the software development process to evaluate the trustworthiness of the produced software~\cite{xiaoyan2012evidence}. 
The use of the word evidence is to have a common abstraction for an argument of a claim with different technologies~\cite{bontekoe2023verifiable,castillo2025trusted}, rather than having a unique perfect mathematical proof concept~\cite{russinovich2024confidential}. 

The evidence life cycle consists of four main stages: Raw, Authenticated, Attested, and Actioned as described in TrustOps~\cite{britotrustops}. 
The life cycle of evidence within a CI pipeline is critical to ensuring the trustworthiness of the software life cycle, which has a continuous evolution and is key to providing verifiable and tamper-resistant trustworthy software artifacts. 
Each action taken using evidence generates new evidence that feeds back into the cycle to continuously enhance trust.
Based on~\cite{britotrustops}, we specify the evidence life cycle as follows:

\textbf{Raw Evidence ($E_{Raw}$):} This stage involves the initial collection of data and logs generated throughout the phases in the CI pipeline. This includes code commits, configuration changes, build logs, test outputs, and metadata such as timestamps, contributor identities, and system environment states. %

\textbf{Authenticated Evidence ($e_{auth}$):} Once raw evidence is collected, it is authenticated by adding a verification mechanism to its source. Authentication involves mechanisms such as developer-signed commits or cryptographically signed build artifacts. This allows commits to be signed by developers, or to ensure that build artifacts are generated in secure environments such as TEEs.%

\textbf{Attested Evidence ($e_{att}$):} After authentication, evidence is subjected to attestation, where its validity is verified in a broader context. Attested evidence builds on authenticated evidence by proving that specific requirements or standards have been met. For example, it means attesting that a build artifact was produced using verified source code and that the test results were generated in a secure, isolated environment. Attestation ensures that evidence can be trusted not only locally, but also by external verifiers, creating a chain of trust throughout the software life cycle.

\textbf{Actioned Evidence ($e_{act}$):} Finally, actioned evidence refers to evidence that directly influenced decision-making processes or triggered automated actions within the CI pipeline. For instance, only attested test results might trigger the deployment of the software to a production environment or revoke access if a security policy is violated. In this context, actioned evidence represents evidence that was used to ensure that only trusted and verified artifacts progressed, enabling continuous trust and accountability in distributed CI pipelines.

\subsection{Protocol using Trustworthiness Mechanisms} \label{sec:evidence_trustworthiness}

The protocol's trustworthiness mechanisms (M1, M2, M3) operate in a strict choreography where each task iteration is executed within a TEE via DBS ($\mathbb{D}$). This ensures that even the policy evaluation ($f_{eval}$) remains tamper-proof and reproducible. By "actioning" the prior state $e_{att}^{i-1}$ and binding outputs through a cryptographic hash $(\parallel \mathcal{H}(e_{act}^{i-1}))$, the protocol produces an immutable DAG. This enables post-facto auditing of the end-to-end provenance by tracing blockchain-anchored commitments ($c_{ledger}$) back to the initial trigger. The process follows~\Cref{alg:evidence_lifecycle}:

\begin{algorithm}
\caption{Evidence-driven Protocol for Trustworthy CI Pipeline}
\label{alg:evidence_lifecycle}
\begin{algorithmic}[1]
\REQUIRE Trigger $E_{Raw}^0$, Tasks $T_{1..n}$, Keys $K$, Ref. Environment $Env_{ref}$, Policies $\Pi$.
\ENSURE Final commitment $c_{ledger}^n$ and actioned evidence $e_{act}^n$, or $\bot$.

\STATE \textbf{Init:} $e_{auth}^0 \leftarrow f_{auth}(E_{Raw}^0, K_{actor})$; \textbf{if} $e_{auth}^0 == \bot$ \textbf{return} $\bot$
\STATE \quad $e_{att}^0 \leftarrow e_{auth}^0$; $c_{ledger}^0 \leftarrow f_{commit}(e_{att}^0)$ \COMMENT{Baseline origin state}

\FOR{$i = 1 \dots n$ \textbf{executing within TEE via DBS ($\mathbb{D}$)}}
    \STATE \textbf{Action:} $(a^i, e_{act}^{i-1}) \leftarrow f_{eval}(e_{att}^{i-1}, \Pi)$ \COMMENT{Policy evaluation}
    \STATE \quad \textbf{if} $a^i \in \{\text{REJECT}, \text{REVOKE}\}$ \textbf{return} $(\text{ABORT: Stage } i \text{ Policy}, \bot)$
    
    \STATE \textbf{Raw:} $E_{Raw}^i \leftarrow \mathbb{D}(t_i, e_{act}^{i-1})$ \COMMENT{Reproducible task execution}
    \STATE \textbf{Auth:} $e_{auth}^i \leftarrow f_{auth}(E_{Raw}^i \parallel \mathcal{H}(e_{act}^{i-1}), K_{system})$ \COMMENT{DAG Binding}
    \STATE \quad \textbf{if} $e_{auth}^i == \bot$ \textbf{return} $\bot$
    
    \STATE \textbf{Attest:} $e_{att}^i \leftarrow f_{attest}(e_{auth}^i, Env_{ref})$; \textbf{if} $e_{att}^i == \bot$ \textbf{return} $\bot$
    \STATE \textbf{Ledger:} $c_{ledger}^i \leftarrow f_{commit}(e_{att}^i)$ \COMMENT{Anchor step to blockchain}
\ENDFOR

\STATE \textbf{Deploy:} $(a^{final}, e_{act}^n) \leftarrow f_{eval}(e_{att}^n, \Pi)$ \COMMENT{Final policy check}
\STATE \quad \textbf{if} $a^{final} \notin \{\text{REJECT}, \text{REVOKE}\}$ \textbf{then} $f_{feedback}(c_{ledger}^n, e_{act}^n)$
\RETURN $(c_{ledger}^n, e_{act}^n)$
\end{algorithmic}
\end{algorithm}

By following the state transitions defined in Algorithm 1, the protocol ensures that every artifact is backed by a verifiable chain of custody. This formal progression directly addresses the identified threat scenarios by preventing the advancement of any task that fails integrity checks.

To mitigate compromised pipeline processes (\textbf{S1}), the protocol secures the execution environment and source origins. %
During execution, secure workflow orchestration (\textbf{M1}) isolates tasks within hardware-protected TEEs, verifying the authenticity of commits and source code modifications (\textbf{TR1}), and ensures that only verified identities can sign or attest to the outputs, protecting the integrity of the processes and environments (\textbf{TR2, TR4}). %

To address risks introduced by third-party dependencies (\textbf{S2}), the system enforces strict validation and reproducible environments. Secure workflow orchestration (\textbf{M1}) requires validation reports for external components before the pipeline can progress, isolating these checks within TEEs. Deterministic build systems (\textbf{M2}) further neutralize dependency risks by ensuring builds are reproducible. This combination guarantees that artifacts remain consistent and verifiable (\textbf{TR4}), even when incorporating external code.

To counter the insufficiency of continuous evidence and audit trails (\textbf{S3}), the protocol establishes a verifiable chain of custody. Secure workflow orchestration (\textbf{M1}) structures and schedules tasks to ensure evidence is systematically generated at each pipeline step. The artifact registry (\textbf{M3}) then binds this data, creating a tamper-proof chain of evidence that securely links final artifacts back to their source code and build processes. This fulfills process and artifact integrity requirements (\textbf{TR3, TR4}) by ensuring provenance is maintained and fully auditable throughout the continuous integration pipeline.

Consequently, this verifiable chain of evidence, materializing the TrustOps evidence lifecycle~\cite{britotrustops}, enables lightweight artifact verification. Because the entire evidence DAG is cryptographically bound (with evidence produced by TEEs running DBS tasks and attested during the pipeline's execution phase), stakeholders are no longer required to perform redundant, independent source rebuilds to establish trust. Instead, artifact verification is reduced to constant-time ($O(1)$) cryptographic signature and policy checks against the final attested evidence.

\section{Protocol Proof of Concept Implementation}\label{ch:implementation}
\begin{figure}[ht]
    \centering
    \includegraphics[width=0.7\columnwidth]{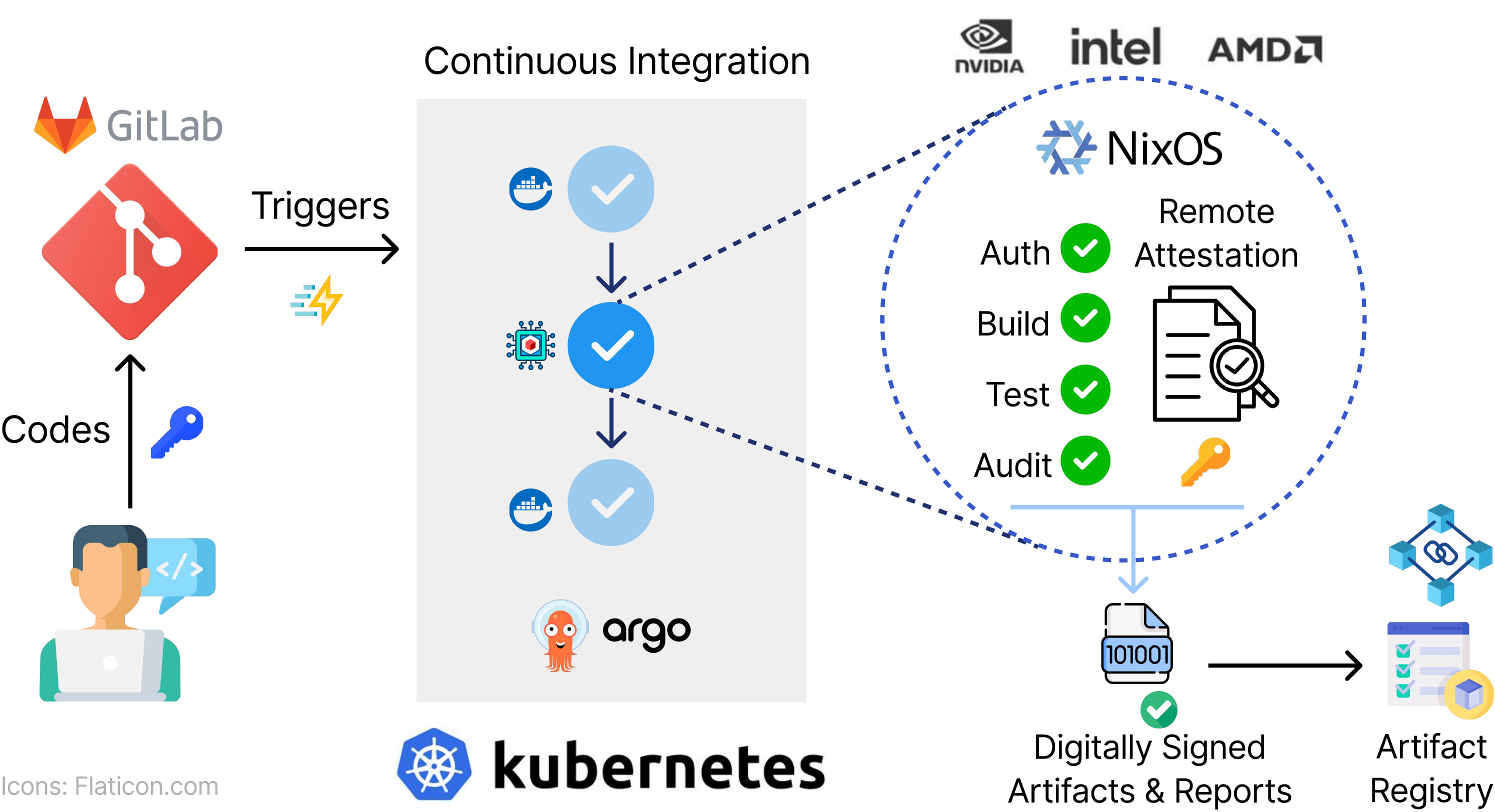}
    \caption{Evidence-driven Trustworthy CI Pipeline Implementation.}
    \label{fig:trustworthy-ci-pipeline-implementation}
\end{figure}

To demonstrate practical feasibility, we prototyped an example of a Trustworthy CI pipeline implementation, available in our public repository\footnote{Repository: \url{https://github.com/trustops/trustworthy-ci-pipelines/}}. Our implementation is a first instantiation of the TrustOps methodology for CI environments~\cite{britotrustops}. It follows the presented protocol and leverages a combination of technologies to enforce security, verifiability, and reproducibility throughout the software development life cycle. %
Figure~\ref{fig:trustworthy-ci-pipeline-implementation} illustrates the implementation.

\subsection{Workflow Orchestration}
\begin{figure*}[ht]
    \centering
    \includegraphics[width=0.80\textwidth]{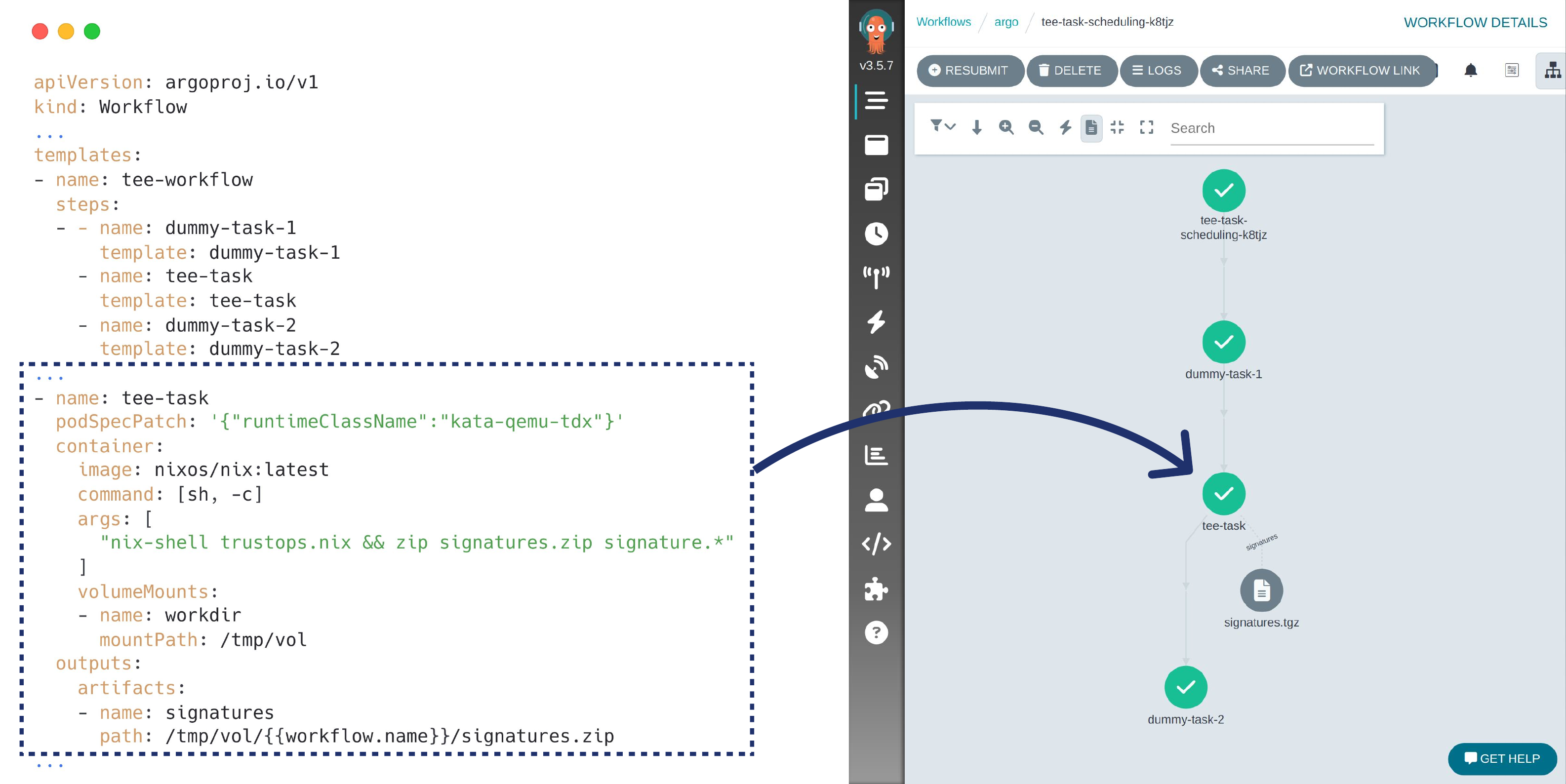}
    \caption{Argo Workflow example pipeline, as YAML file on the left, comprising one TEE-based task, labeled via the $runtimeClassName$ attribute as running inside an Intel TDX VM, and its execution and visualization on the right.}
    \label{fig:argo-workflows-tee-orchestration}
\end{figure*}

The pipeline is built around GitLab, which serves as the central VCS. The pipeline execution is triggered by events like commit pushes or merge requests. Every commit is authenticated through developer-signed commits, ensuring that only trusted identities can introduce changes. %
Argo Workflows is employed for orchestrating the CI tasks, operationalizing the CI pipeline in sequence according to dependency graphs defined in pipeline configurations. It runs on top of a Kubernetes cluster, which provides scalability and resource management. %
Tasks, defined by the developers in the pipeline specification, are executed in TEEs, and the Argo engine handles the task orchestration within the Kubernetes cluster. Before execution, each TEE should perform remote attestation to verify it is operating on trusted hardware with an untampered software stack, and exchanges trusted signing keys to establish a secure environment. These tasks have specific labels that trigger their placement and verifiable execution. An illustration of the labeling and pipeline orchestration is shown in Figure~\ref{fig:argo-workflows-tee-orchestration}.

\subsection{Declarative and Reproducible Builds}

The build process relies on NixOS and Nix, which enforce a declarative, reproducible, and deterministic build environment. It allows to capture the full dependency tree and build instructions in a declarative format, ensuring that every build is reproducible, enabling any stakeholder to independently verify the results. The Nix script running inside the TEE VM authenticates the source code and performs builds and tests in an isolated, reproducible, and verifiable environment. This ensures that the same inputs (e.g., code, dependencies) always produce the same outputs, regardless of when or where the build is performed.

\subsection{Testing, Auditing, and Output Signing}

Although flexible, tests can also run inside the same reproducible, isolated environment to ensure the integrity of the testing process. Audit logs are generated and signed, during both build and testing phases, or as part of additional audit processes, such as security audits, creating a transparent and traceable record of authenticated evidence for each task. Upon successful execution, the build artifacts, test results, and audit logs are digitally signed using the TEE's trusted cryptographic keys. These evidence outputs are then uploaded to a secure artifact storage system, ensuring their integrity and providing verifiable proof that they were generated by trusted processes.

\subsection{Artifact Storage and Verifiable Evidence}

The final step in the pipeline is the secure storage of artifacts in an artifact registry. The artifacts and reports, now digitally signed, are uploaded with cryptographic proof of their provenance. These attestations are then committed into a blockchain, via smart contracts, using the Ethereum Attestation Service\footnote{https://docs.attest.org} standard for efficient off-chain verifications~\cite{boi2024ethereum}, enabling the creation of a tamper-proof chain of evidence, where all actions within the CI pipeline are linked through cryptographic proofs. This approach enables post facto verification of specific policies, allowing retrospective confirmation at each phase in the CI pipeline. %
This allows external auditors and stakeholders to independently verify the integrity, provenance, and assurance of any artifact produced by the CI pipeline.

\section{Protocol Evaluation}\label{ch:evaluation}
To evaluate the properties our system, we modeled a scenario involving trust establishment between a Provider and a Consumer within a software supply chain. %
The Provider represents or controls \textbf{E1} to \textbf{E6}, while the Consumer controls \textbf{E7}. %
This scenario can be generalized to various contexts, including intra-organization dependencies, business-to-business software supply chains, and open-source software.

We modeled this scenario following our implementation of the Evidence-driven Trustworthy CI pipeline protocol. For testing, to represent the 3 threat scenarios (\textbf{S}), we used a frontend interface\footnote{https://github.com/aave/interface}, a backend service\footnote{https://github.com/ChainSafe/ChainBridge/}, and an oracle application\footnote{https://github.com/smartcontractkit/chainlink}, each with a suite of tests and known vulnerabilities, detectable by tools such as trivy\footnote{https://github.com/aquasecurity/trivy}. We compared two variations of the scenario, using the same pipeline, one using the mechanisms for trustworthiness and one without them.%

\subsection{Results and Threat Mitigation}

Our results, summarized in Table~\ref{tab:results}, demonstrate that the system introduces a manageable overhead while significantly enhancing evidence quality and reducing verification complexity for the Consumer. %

\begin{table*}[!ht]
\tiny
\setlength{\tabcolsep}{3pt}
\renewcommand{\arraystretch}{1.5}
\centering
\caption{Comparison Results of 10 runs across Three Use Cases}
\label{tab:results}
\begin{tabular}{|p{2.5cm}|c|c|c|c|c|c|}
\hline
\multirow{2}{*}{\textbf{Metric}} & \multicolumn{2}{c|}{\textbf{Oracle Core}} & \multicolumn{2}{c|}{\textbf{Frontend Interface}} & \multicolumn{2}{c|}{\textbf{Backend Service}} \\ \cline{2-7}
 & \textbf{Without} & \textbf{With} & \textbf{Without} & \textbf{With} & \textbf{Without} & \textbf{With} \\ \hline
Total Workflow Time (min) &
$3.32 \pm 0.13$ & $6.00 \pm 0.24$ &
$4.98 \pm 0.20$ & $9.00 \pm 0.36$ &
$9.13 \pm 0.37$ & $16.50 \pm 0.66$ \\ \hline
CPU Consumption (s / CPU) &
$21.60 \pm 0.86$ & $39.20 \pm 1.57$ &
$32.40 \pm 1.30$ & $58.80 \pm 2.35$ &
$59.40 \pm 2.38$ & $107.80 \pm 4.31$ \\ \hline
Memory Consumption (s / 100 MiB) &
$4.80 \pm 0.19$ & $9.60 \pm 0.38$ &
$7.20 \pm 0.29$ & $14.40 \pm 0.58$ &
$13.20 \pm 0.53$ & $26.40 \pm 1.06$ \\ \hline
Runtime of Tasks (s) &
$52.64 \pm 2.11$ & $113.20 \pm 4.53$ &
$78.96 \pm 3.16$ & $169.80 \pm 6.79$ &
$144.76 \pm 5.79$ & $311.30 \pm 12.45$ \\ \hline
Evidence Size (B / signature) &
N/A & $294 \pm 0$ &
N/A & $294 \pm 0$ &
N/A & $294 \pm 0$ \\ \hline
Cryptographic Operation Time (ms / Op) &
N/A & $80 \pm 3$ &
N/A & $80 \pm 2$ &
N/A & $80 \pm 2$ \\ \hline
Blockchain Commit Cost (gas)            & N/A           & 393k$\pm$11k    & N/A           & 393k$\pm$11k   & N/A           & 393k$\pm$11k    \\ \hline
Blockchain Revocation Cost (gas)            & N/A           & 6.4k   & N/A           & 88k   & N/A           & 88k   \\ \hline
\end{tabular}
\end{table*}

These numbers also depend on factors such as the size of the codebase, the complexity of the build process, and the volume of generated artifacts. Despite this, the results show that the performance remains within the same order of magnitude, for the comparison between the two workflows, with and without mechanisms across the three codebases. The TEE-based workflow offers considerable potential for optimization in areas such as orchestration, image management, and networking overhead, especially with better integration into the workflow engine. %
Additionally, TEE hardware is known to introduce another layer of overhead~\cite{kumar2020performance}.

\begin{figure}[h]
    \centering
    \includegraphics[width=0.80\columnwidth]{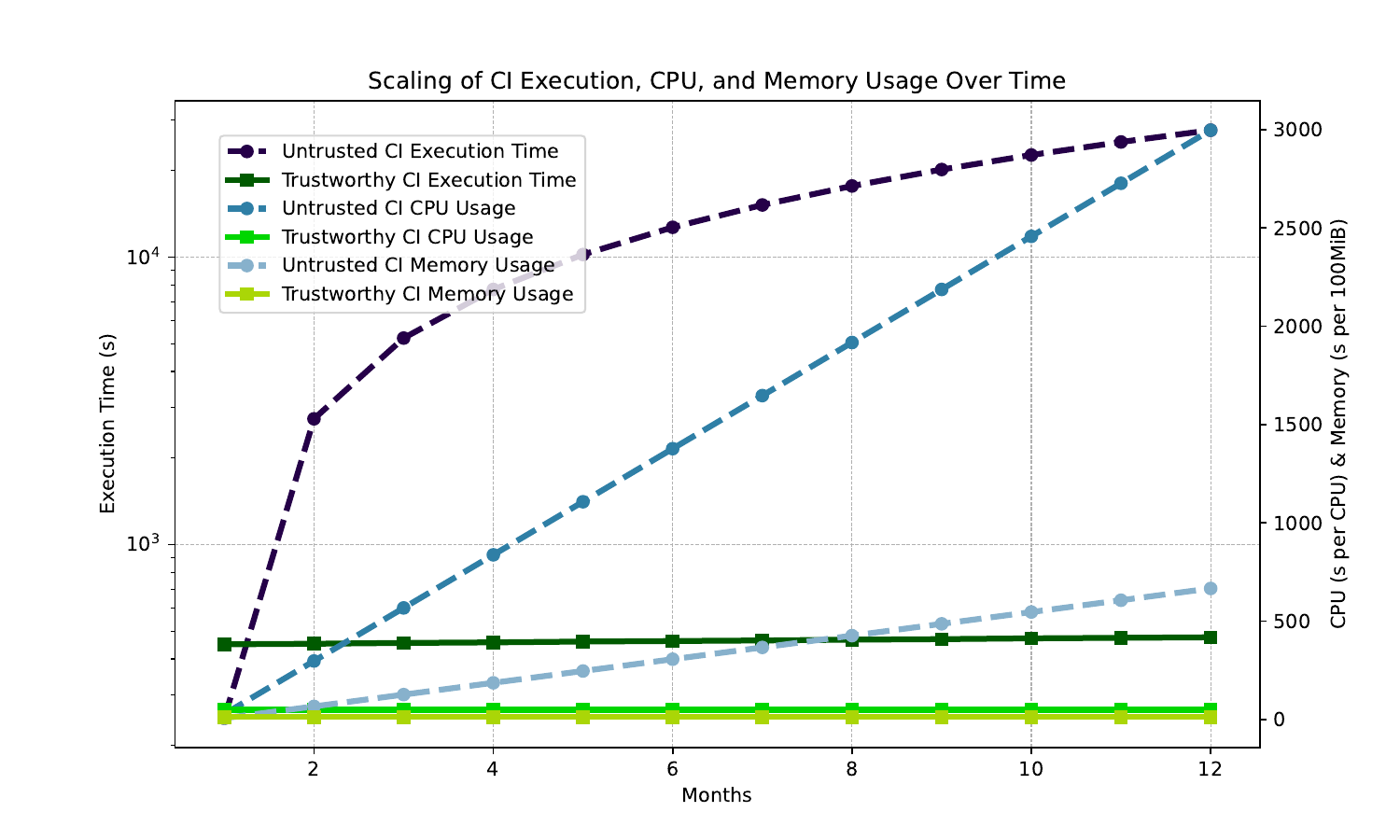}
    \caption{Scaling of CI execution, CPU, and memory usage over time, for a growth rate of 10 consumers per month and a new release every month. The nature and size of the code base remained approximately constant. The Execution times have their scale on the left axis, while CPU and Memory usage have their scale on the right axis.}
    \label{fig:trustworthy-ci-scaling}
\end{figure}

By cryptographically binding CI outputs to TEE-attested execution, Consumers can efficiently verify artifact integrity, authenticity, and policy compliance using lightweight checks.
To quantify this effect, we simulated a software release process in which the number of Consumers grows by a factor of 10 per month\footnote{Many open-source projects may experience even larger user growth rates and more dynamic release cycles~\cite{borges2018s}}, while each release remains approximately constant in size and complexity, representing development patterns such as iterative bug fixes, feature adjustments, or controlled modifications with balanced additions and deletions~\cite{deshpande2008total}. In the untrusted CI model, every Consumer independently rebuilds, tests, and audits each release. In contrast, under our Trustworthy CI Protocol, CI execution is performed once per release, and each Consumer performs only cryptographic verification. As shown in Figure~\ref{fig:trustworthy-ci-scaling}, the cumulative execution cost in the untrusted model grows rapidly with the number of Consumers, whereas the marginal overhead at the Producer’s pipeline in our approach remains nearly constant. This demonstrates that verification costs are effectively amortized across Consumers, enabling scalable trust in highly collaborative software supply chains where artifacts are shared across multiple stakeholders. 

To validate threat mitigation, we simulated \textbf{S1} by attempting to modify a build script after TEE attestation, resulting in a failed artifact signature verification, and \textbf{S2} vulnerabilities from third-party dependencies detected with the SBOM generation. For \textbf{S3}, we altered a committed hash in the artifact registry; the registry verification detected the inconsistency within $\leq$100ms.

\section{Discussion}\label{ch:discussion}
There are several points to consider when using a trustworthy CI protocol.
A major weakness of TEEs is the reliance on the hardware manufacturer which controls the public key infrastructure (PKI) for the TEE identity keys. Furthermore, several attack scenarios for TEE exist, including memory corruption, as demonstrated in~\cite{tee-fail}. 
An important factor of the security and practical effectiveness of TEEs and blockchain commitments relies heavily on the accuracy and integrity of their formal verification to guarantee their desired properties~\cite{sabt2015trusted}. Additionally, it is important to consider that evidence reproducibility can be affected by non-stable inputs, such as timestamps or environment differences, which prevent full determinism in the build process~\cite{poll2022automating}.
Incorrectly defined constraints could lead to evidence that is technically valid but fail to represent the intended computation accurately.
Undetected vulnerabilities or malicious code could compromise the computation's integrity, resulting in incorrect or insecure outcomes, even within the secure confines of a TEE. %

One additional strength of our CI protocol is its seamless integration into existing CI/CD pipeline tools. We show that it is possible to adapt and leverage widely adopted platforms such as GitLab, Argo Workflows, and Kubernetes to ensure minimal disruption to current development practices~\cite{singh2022deploying}. This approach eliminates the need for extensive custom development or reconfiguration, making the adoption process straightforward and accessible, reducing barriers to entry, while maintaining scalability and performance.
Despite challenges like computational overhead and complexity, blockchain remains a practical solution for improving trust in the evidence verification and validation process in software supply chains~\cite{helliar2020permissionless}.

Our evaluation showed that our mechanisms can help establish trust in software supply chains and  the results demonstrate the viability of this approach. While our implementation introduces modest overhead during release preparation, with no impact in day-to-day development, it significantly reduces verification complexity for Consumers, especially in collaborative environments. %
However, at organizational scale, additional factors warrant consideration: (1) concurrent pipeline scheduling may create contention for TEE-enabled nodes, requiring capacity planning or queuing strategies; (2) blockchain commitment costs could be reduced through batching attestations across multiple pipeline runs; (3) attestation storage will grow over time, potentially benefiting from off-chain storage with on-chain anchors. Although the modular design of our protocol supports horizontal scaling of the container orchestration layer, a systematic evaluation of large-scale deployment patterns is left to future work. An interesting direction for future work is the integration of build-time integrity guarantees with mechanisms for verifying software composition after artifact generation. Combining TEE-based attestation with confidentiality-preserving SBOM verification~\cite{trustbom_2026} could enable end-to-end verifiable supply chains spanning artifact creation through deployment.

\section{Conclusion}\label{ch:conclusion}

In this paper, we introduced an evidence-driven protocol that secures continuous integration pipelines by combining DBS with TEEs. This framework generates immutable, blockchain-backed attestations to verify artifact authenticity, integrity, and provenance throughout the software life cycle. Our evaluation across three software use cases demonstrates that this approach mitigates specific supply chain threats with a measurable and manageable computational overhead. By shifting the burden of trust establishment to the pipeline's execution phase, the protocol reduces consumer-side validation to lightweight signature checks. This eliminates the need for redundant, independent rebuilds by consumers, effectively amortizing the initial TEE overhead in multi-consumer scenarios. These results indicate that the proposed framework offers a viable and verifiable trust model for software infrastructure.

\bibliographystyle{src/splncs04}
\makeatletter
\if@filesw
\immediate\write\@auxout{\string\bibdata{_refs}}%
\fi
\makeatother

\end{document}